\documentclass[aps,twocolumn,prl]{revtex4-1}
\usepackage{graphicx,amsmath, amsfonts,bm, color}

\usepackage{xcolor,hyperref}
\hypersetup{
   colorlinks,
   linkcolor={blue!50!black},
   citecolor={blue!50!black},
   urlcolor={blue!80!black}
}

\usepackage{enumitem}

\renewcommand{\=}{\!=\!}
\newcommand{\1}{^{\mbox{\tiny (1)}}}

\newcommand{\tr}{\operatorname{tr}}

\begin{document}

\title{Universality and stability phase-diagram of two-dimensional brittle fracture}
\author{Yuri Lubomirsky$^{1}$}
\thanks{Y.~Lubomirsky and C-H.~Chen contributed equally.}
\author{Chih-Hung Chen$^{2}$}
\thanks{Y.~Lubomirsky and C-H.~Chen contributed equally.}
\author{Alain Karma$^{2}$}
\thanks{eran.bouchbinder@weizmann.ac.il (E.~Bouchbinder), a.karma@northeastern.edu (A.~Karma).}
\author{Eran Bouchbinder$^{1}$}
\thanks{eran.bouchbinder@weizmann.ac.il (E.~Bouchbinder), a.karma@northeastern.edu (A.~Karma).}


\affiliation{$^{1}$Chemical and Biological Physics Department, Weizmann Institute of Science, Rehovot 7610001, Israel\\
$^{2}$Department of Physics and Center for Interdisciplinary Research on Complex Systems, Northeastern University, Boston, Massachusetts 02115, USA}

\begin{abstract}
The two-dimensional oscillatory crack instability, experimentally observed in a class of brittle materials under
strongly dynamic conditions, has been recently reproduced by a nonlinear phase-field fracture
theory. Here we highlight the universal character of this instability by showing that it is present
in materials exhibiting widely different near crack tip elastic nonlinearity, and by demonstrating that
the oscillations wavelength follows a universal master curve in terms of dissipation-related and
nonlinear elastic intrinsic length scales. Moreover, we show that upon increasing the driving force for
fracture, a high-velocity tip-splitting instability emerges, as experimentally demonstrated. The analysis
culminates in a comprehensive stability phase-diagram of two-dimensional brittle fracture, whose salient
properties and topology are independent of the form of near tip nonlinearity.
\end{abstract}

\maketitle

Cracks mediate materials failure and hence understanding their spatiotemporal dynamics is of prime fundamental and practical importance~\cite{freund1998dynamic, Broberg.99, Bouchbinder.10, Bouchbinder.14, Fineberg.99}. It is experimentally well-established that cracks in brittle materials undergo various symmetry-breaking instabilities~\cite{Fineberg.99, Ravi-Chandar.84c, Fineberg.91, Sharon.96, livne2005universality, boue2015origin, kolvin2018topological, Livne.07, Goldman.12}. Most notably, straight cracks in three-dimensional (3D) samples of isotropic materials universally undergo a micro-branching instability, where short-lived micro-cracks branch out sideways from the main crack, at low to medium propagation velocities~\cite{Fineberg.99, Ravi-Chandar.84c, Fineberg.91, Sharon.96, livne2005universality, boue2015origin, kolvin2018topological}. Recent experiments performed on a class of neo-Hookean (\mbox{\footnotesize NH}) --- a nonlinear extension of linear, Hookean elasticity --- brittle materials revealed that upon reducing the system's thickness, approaching the two-dimensional (2D) limit, the micro-branching instability is severely suppressed; as a result, straight cracks accelerate to unprecedentedly high velocities, approaching the relevant sonic velocity, until they undergo an oscillatory instability~\cite{Bouchbinder.14,Livne.07, Goldman.12}.
\begin{figure}[ht!]
 \centering
 \includegraphics[width=0.48\textwidth]{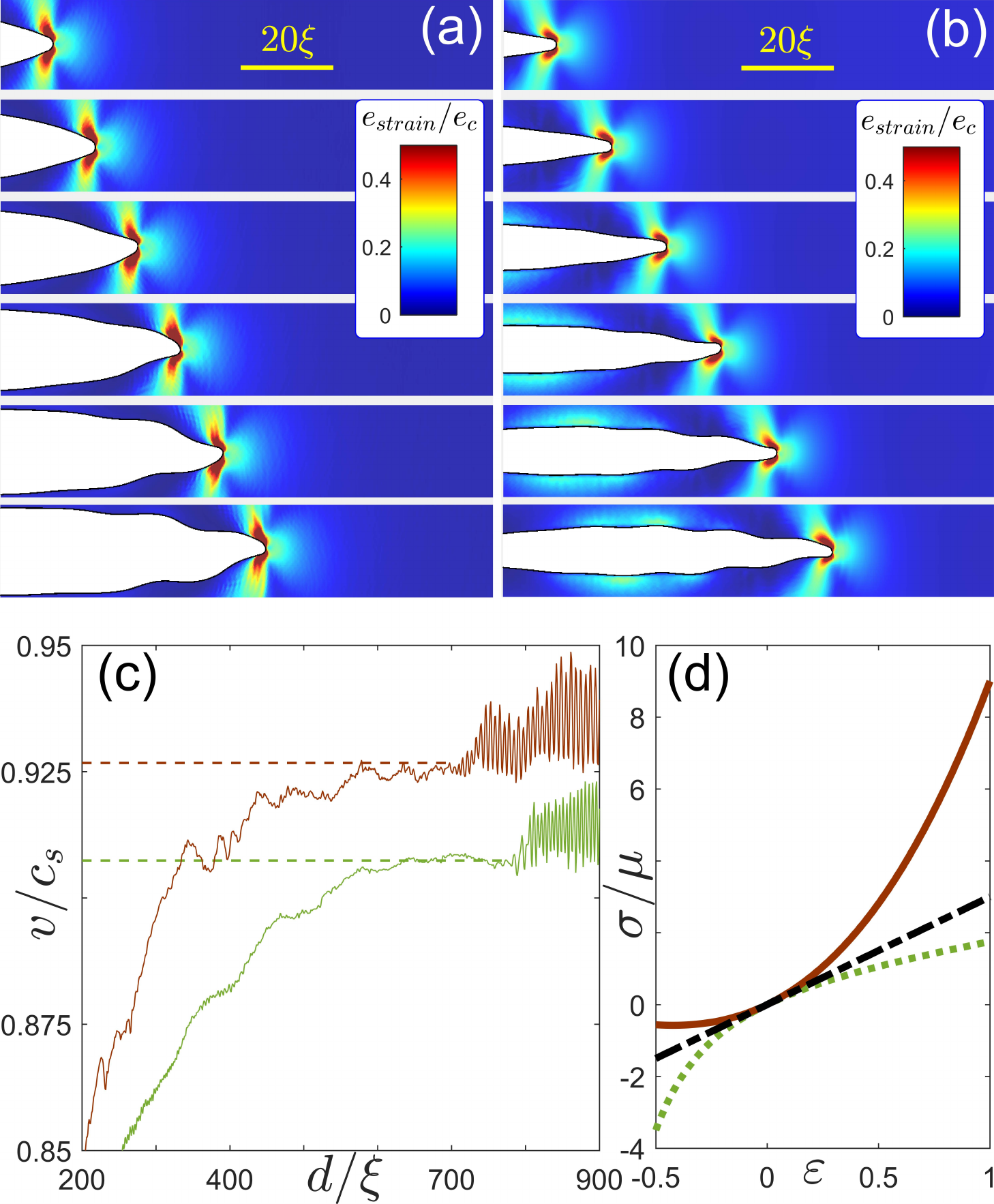}
 \caption{(a) A zoom in on the 2D oscillatory instability in a phase-field simulation of brittle neo-Hookean (\mbox{\footnotesize NH}) materials under tensile (mode I) loading~\cite{Chen:2017aa}. The color code corresponds to the normalized strain energy density and $\xi$ is the dissipation length. (b) The same, but for brittle Saint-Venant-Kirchhoff (\mbox{\footnotesize SVK}) materials, see text and~\cite{SM} for details. (c) Examples of the normalized crack speed $v/c_s$ ($c_s$ is the shear wave speed) vs.~normalized crack propagation distance $d/\xi$ for \mbox{\footnotesize SVK} (top curve) and \mbox{\footnotesize NH} materials (bottom curve). The onset of oscillations is marked by the horizontal dashed lines. (d) The uniaxial stress $\sigma$ (normalized by the shear modulus $\mu$) vs.~strain $\varepsilon$ for \mbox{\footnotesize SVK} (solid line) and \mbox{\footnotesize NH} materials (dashed line), along with their linear approximation (dashed-dotted line).}
 \label{fig:oscillations}
\end{figure}

Our understanding of these dynamic fracture instabilities is far from complete. In particular, the classical theory of brittle cracks --- Linear Elastic Fracture Mechanics (LEFM)~\cite{freund1998dynamic, Broberg.99} --- intrinsically falls short of explaining these instabilities~\cite{Bouchbinder.14,Bouchbinder.09b, Chen:2017aa}. Recently, integrating theoretical ideas about the existence and importance of elastic nonlinearity near cracks tips~\cite{Livne.08, Bouchbinder.08,Bouchbinder.09,livne2010near,Goldman.12, Bouchbinder.14} into a phase-field fracture theory, the 2D oscillatory instability in brittle \mbox{\footnotesize NH} materials has been quantitatively reproduced in large-scale simulations for the first time~\cite{Chen:2017aa}, cf.~Fig.~\ref{fig:oscillations}a. The theory has shown, in agreement with experiments, that the wavelength of oscillations scales linearly with an intrinsic nonlinear elastic length scale, which does not exist in LEFM, hence explicitly demonstrating the failure of the classical theory of brittle cracks.

This significant progress raised several pressing questions of fundamental importance. First, is the oscillatory instability universal, i.e.~observed in various brittle materials independently of the nature and form of near tip elastic nonlinearity? Second, what are the minimal physical conditions for the existence of the oscillatory instability? Third, are there additional, previously undiscovered instabilities in 2D dynamic fracture? Finally, can one derive a comprehensive stability phase-diagram of two-dimensional brittle fracture and if so, is it universal? In this Letter, we extensively address these important questions using theoretical considerations and large-scale simulations in the framework of the recently developed nonlinear phase-field theory of fracture~\cite{Chen:2017aa}.

This theory belongs to a broader class of diffuse-interface approaches to fracture~\cite{bourdin2000numerical,Karma.01,Hakim.09,pons2010helical,spatschek2011phase,bourdin2014morphogenesis} that avoid the difficulty of tracking the evolution of sharp crack surfaces, and at the same time allow a self-consistent selection of the crack's velocity and path, which are far from being understood in general. These properties emerge from the dynamics of an auxiliary phase-field $\phi$ and its coupling to other fields, which provide a mathematical machinery that renders the fracture problem self-contained. In particular, it gives rise to a dissipation-related length scale $\xi$ and a fracture energy $\Gamma(v)$ (the energy dissipated per unit crack surface, where $v$ is the crack propagation). The smooth transition in space between the pristine ($\phi\!=\!1$) and the fully broken ($\phi\!=\!0$) states of the material is initiated when the elastic strain energy functional $e_{\mbox{\tiny{strain}}}(\bm H)$ exceeds a threshold $e_c$, where $\bm H\=\nabla{\bm u}$ is the displacement gradient tensor and ${\bm u}(x,y,t)$ is the displacement vector field.

Unlike previous phase-field approaches~\cite{Karma.01,Hakim.09, spatschek2011phase}, the formulation in~\cite{Chen:2017aa} maintains the wave speeds constant inside the dissipation zone, thereby allowing cracks to accelerate to unprecedentedly high velocities without undergoing tip-splitting instabilities at marginally dynamic velocities~\cite{Karma2004, Henry.08}. Moreover, $e_{\mbox{\tiny{strain}}}(\bm H)$ is taken to be nonlinear, introducing nonlinearity on a length $\ell$ in the near vicinity of the crack tip, while the linear elastic (quadratic) approximation to $e_{\mbox{\tiny{strain}}}(\bm H)$ remains very good on larger scales. Hence, this theory features an intrinsic dissipation length $\xi$ and intrinsic length $\ell$ associated with near tip elastic nonlinearity, which can be independently varied.
\begin{figure*}[ht!]
 \centering
 \includegraphics[width=0.95\textwidth]{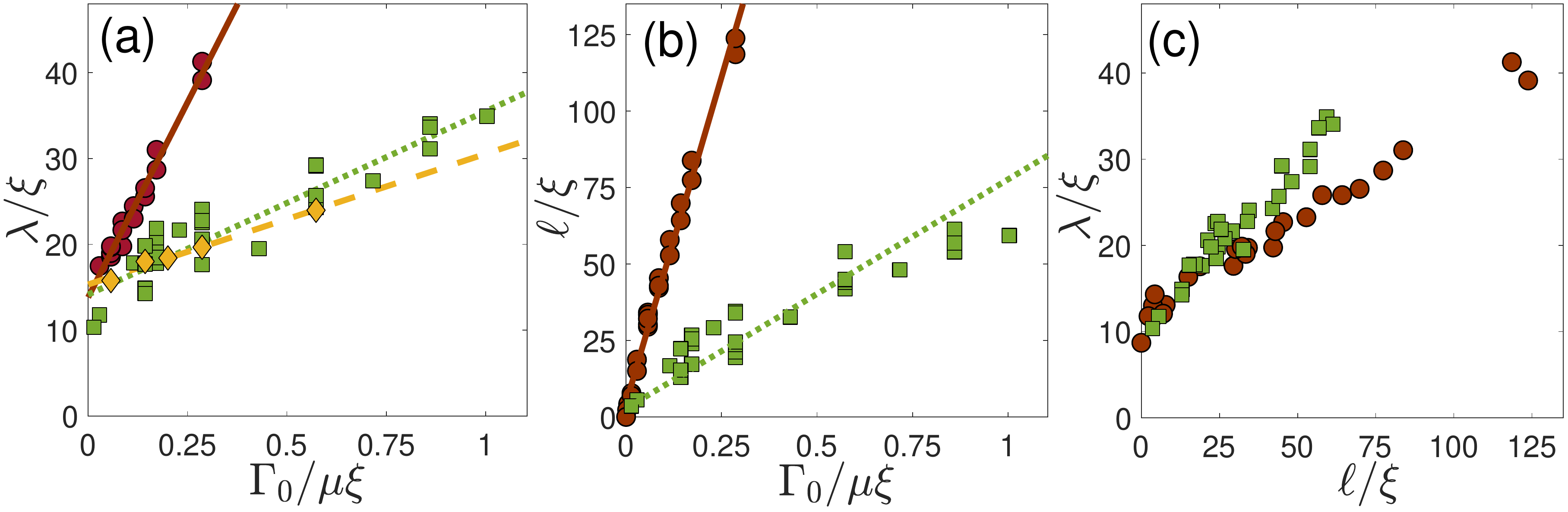}
 \caption{(a) The normalized oscillations wavelength $\lambda/\xi$ vs.~$\Gamma_0/\mu\xi$ for \mbox{\footnotesize SVK} (circles), \mbox{\footnotesize NH} (squares) and truncated \mbox{\footnotesize NH} (diamonds) materials. (b) The normalized nonlinear length $\ell/\xi$ ($\ell$ is defined in the text) vs.~$\Gamma_0/\mu\xi$ for \mbox{\footnotesize SVK} (circles) and \mbox{\footnotesize NH} (squares) materials. (c) $\lambda/\xi$ vs.~$\ell/\xi$ for \mbox{\footnotesize SVK} (circles) and \mbox{\footnotesize NH} (squares) materials. The lines are guides to the eye.}
 \label{fig:wavelength}
\end{figure*}

The strain energy functional $e^{\mbox{\tiny NH}}_{\mbox{\tiny{strain}}}(\bm H)\!=\!\frac{1}{2}\mu(\tr({\bm F}^{\rm T}{\bm F})\!+\![\det({\bm F})]^{-2}\!-\!3)$ of 2D incompressible \mbox{\footnotesize NH} materials~\cite{Knowles.83}, where ${\bm F}\={\bm I}\!+\!{\bm H}$ (${\bm I}$ is the identity tensor) and $\mu$ is the shear modulus, has been shown~\cite{Chen:2017aa} to quantitatively predict the experimentally observed oscillatory instability in 2D brittle polymeric elastomers under tensile (mode I) loading conditions~\cite{Livne.07,Goldman.12, Bouchbinder.14}. Most notably, the oscillations (see example in Fig.~\ref{fig:oscillations}a) emerge at a critical velocity $v_c$ (see the lower horizontal dashed line in Fig.~\ref{fig:oscillations}c) and their wavelength follows $\lambda\!=\!\alpha\,\xi \!+\!\beta\,\Gamma_0/\mu$ (shown by the squares in Fig.~\ref{fig:wavelength}a), both in quantitative agreement with experiments~\cite{Goldman.12,Chen:2017aa}. Here $\alpha$ and $\beta$ are dimensionless numbers, $\xi$ is the dissipation length defined above, $\Gamma_0\!\equiv\!\Gamma(v\!\to\!0)$ and $\Gamma_0/\mu$ is proportional to the intrinsic nonlinear elastic length $\ell$~\cite{Bouchbinder.08, livne2010near,Goldman.12, Bouchbinder.14}. $e^{\mbox{\tiny NH}}_{\mbox{\tiny{strain}}}$ is of entropic origin, which is well-understood in terms of the statistical thermodynamics of cross-linked polymer chains~\cite{flory1943statistical}.

To assess the generality of these results, our first goal is to consider near tip elastic nonlinearity which is sufficiently general, yet of qualitatively different physical origin and emerging properties compared to $e^{\mbox{\tiny NH}}_{\mbox{\tiny{strain}}}$. To that aim, we invoke the minimal elastic nonlinearity associated with the rotational invariance of isotropic materials. That is, we use the rotationally-invariant (Green-Lagrange) metric strain tensor ${\bm E}\=\frac{1}{2}({\bm F}^{\rm T}{\bm F}\!-\!{\bm I})\=\bm \varepsilon+\frac{1}{2}(\bm H^T \bm H)$, instead of its widespread linear approximation $\bm \varepsilon \= \frac{1}{2}(\bm H + \bm H^T)$~\cite{Landau.86}. When combined with constitutive linearity, i.e.~with a quadratic energy functional in which the linearized strain measure $\bm \varepsilon$ is replaced by its nonlinear rotationally-invariant counterpart ${\bm E}$, we obtain
\begin{equation}
e^{\mbox{\tiny SVK}}_{\mbox{\tiny{strain}}}(\bm H)\!=\!\tfrac{1}{2}\tilde\lambda\,\text{tr}^2\!\left(\bm E\right)+\mu\,\text{tr}\!\left(\bm E^{2}\right) \ .
\label{eq:SV_energy}
\end{equation}
This energy functional, corresponding to Saint-Venant-Kirchhoff (\mbox{\footnotesize SVK}) materials~\cite{Holzapfel.00}, is constitutively identical to linearized Hookean elasticity (where $\tilde\lambda$ is the first Lam\'e constant~\cite{Landau.86}), but features geometric nonlinearity embedded inside ${\bm E}$.

While materials typically feature constitutive nonlinearity in addition to geometric nonlinearity, they should {\em at least} feature the latter, and hence Eq.~\eqref{eq:SV_energy} constitutes the minimal possible elastic nonlinearity. $e^{\mbox{\tiny SVK}}_{\mbox{\tiny{strain}}}$ is not only of a qualitatively different physical origin compared to $e^{\mbox{\tiny NH}}_{\mbox{\tiny{strain}}}$, but it also exhibits significantly different properties. In Fig.~\ref{fig:oscillations}d, we plot the uniaxial tension response corresponding to the two functionals, in addition to their linear elastic approximation (the elastic constants are chosen such that the latter is identical for both functionals). We observe that $e^{\mbox{\tiny SVK}}_{\mbox{\tiny{strain}}}$ features a stronger nonlinearity than $e^{\mbox{\tiny NH}}_{\mbox{\tiny{strain}}}$ and is of strain-stiffening nature, while the latter is of strain-softening nature. Do materials described by $e^{\mbox{\tiny SVK}}_{\mbox{\tiny{strain}}}$ experience the same oscillatory instability as those described by $e^{\mbox{\tiny NH}}_{\mbox{\tiny{strain}}}$?

In Fig.~\ref{fig:oscillations}b we present results of a large-scale numerical simulation of the nonlinear phase-field fracture theory discussed above, using $e^{\mbox{\tiny SVK}}_{\mbox{\tiny{strain}}}$ (see simulation details in~\cite{SM}). An oscillatory instability that is strikingly similar to the oscillatory instability shown in Fig.~\ref{fig:oscillations}a for brittle \mbox{\footnotesize NH} materials is observed. The onset of instability, see example in Fig.~\ref{fig:oscillations}c (top curve), takes place at an ultra-high critical velocity $v_c$, whose normalized value is similar to the one of brittle \mbox{\footnotesize NH} materials (bottom curve). The fact that the oscillatory instability exists for widely different forms of near crack tip elastic nonlinearity provides a strong indication in favor of its universal nature. In Fig.~\ref{fig:wavelength}a, we present the oscillations wavelength of brittle \mbox{\footnotesize SVK} materials demonstrating that it scales linearly with $\Gamma_0/\mu$, $\lambda\!=\!\alpha\,\xi \!+\!\beta\,\Gamma_0/\mu$, exactly as it does for brittle \mbox{\footnotesize NH} materials, yet again supporting the universality of the oscillatory instability. Remarkably, while $\beta$ is significantly larger for brittle \mbox{\footnotesize SVK} materials compared to brittle \mbox{\footnotesize NH} materials, $\alpha\!=\!13\pm1$ is essentially identical for the two classes of materials.

To address the apparent independence of $\alpha$ on the form of near tip elastic nonlinearity and its physical implications, we consider yet another energy functional. In selecting the latter, we aim at addressing also a different question: Is the wavelength
significantly affected by strong elastic nonlinearity deep inside the near tip nonlinear elastic zone or mainly by weak elastic nonlinearity, as predicted by the weakly nonlinear theory of fracture~\cite{Bouchbinder.08,Bouchbinder.09b,Bouchbinder.14}? To address these questions, we consider the incompressible \mbox{\footnotesize NH} functional truncated to leading order nonlinearity
\begin{align}
&e^{\mbox{\tiny TNH}}_{\mbox{\tiny{strain}}}(\bm H)/\mu\simeq\text{tr}(\boldsymbol{\varepsilon}^{2})\!+\!\text{tr}(\boldsymbol{\varepsilon})^{2}
\!-\!\tfrac{1}{2}\text{tr}(\boldsymbol{H})^{3}\!-\!\tfrac{3}{2}\text{tr}(\boldsymbol{H})\,\text{tr}(\boldsymbol{H}^{2})\nonumber\\
&-\!\tfrac{1}{8}\text{tr}(\boldsymbol{H})^{4}\!+\!\tfrac{9}{4}\text{tr}(\boldsymbol{H})^{2}\,\text{tr}(\boldsymbol{H}^{2})\!+\!
\tfrac{3}{8}\text{tr}(\boldsymbol{H}^{2})^{2}\!+\!{\cal O}({\bm H}^5) \ .
\label{eq:truncated}
\end{align}
Unlike the weakly nonlinear theory of fracture, which is an analytic perturbative theory that takes into account only cubic nonlinearity in the energy, here we are looking for global numerical solutions and hence should include also quartic nonlinearity to ensure the well-posedness of the global fracture problem~\cite{SM}.

Phase-field theory calculations with $e^{\mbox{\tiny TNH}}_{\mbox{\tiny{strain}}}$ of Eq.~\eqref{eq:truncated} demonstrated the existence of an oscillatory instability (not shown), whose wavelength is essentially indistinguishable from that of $e^{\mbox{\tiny NH}}_{\mbox{\tiny{strain}}}$, as shown in Fig.~\ref{fig:wavelength}a (diamonds). This result indicates that weak elastic nonlinearity controls the oscillatory instability and further highlights the independence of $\alpha$ on the form of nonlinearity. The latter raises an intriguing possibility; if elastic nonlinearity vanishes, $\ell\!\propto\!\Gamma_0/\mu\!\to\!0$, the results of Fig.~\ref{fig:wavelength}a imply that $\lambda\!\approx\!13\xi$. That is, the oscillatory instability may exist in the absence of elastic nonlinearity, where its wavelength is determined by the intrinsic dissipation length $\xi$.

This possibility is quite intriguing because, if true, it means that near tip elastic nonlinearity is not necessary for the existence of the oscillatory instability, which can alternatively inherit its characteristic length from the dissipation zone. The extensive calculations that gave rise to the wavelength $\lambda$ in Fig.~\ref{fig:wavelength}a, where the ratio $\Gamma_0/\mu$ has been reduced as much as possible within numerical limitations, also showed that the {\em amplitude} of the oscillations diminishes as $\ell\!\propto\!\Gamma_0/\mu$ is reduced. This observation might support the possibility that the instability exists with a finite wavelength and a vanishingly small amplitude as elastic nonlinearity vanishes. Hence, the observation in~\cite{Chen:2017aa} in which no oscillatory instability is observed when $\ell\!=\!0$, i.e.~using the linear elastic (quadratic) approximation of $e_{\mbox{\tiny{strain}}}$ to begin with, may simply be explained in terms of a vanishing amplitude, not as indicating the absence of instability in this limit. On the other hand, based on presently available evidence, we cannot exclude the possibility that the instability disappears at a finite (yet very small) value of $\ell\!\propto\!\Gamma_0/\mu$, as further discussed below.

While $\alpha$ appears to be independent of the form of near tip elastic nonlinearity, the slope $\beta$ in the linear $\lambda\,$--$\,\Gamma_0/\mu$ relation does depend on it. To understand this dependence, recall that the nonlinear length $\ell$ is proportional to $\Gamma_0/\mu$, but is not identical to it (e.g.~$\Gamma_0/\mu$ exists also in the absence of elastic nonlinearity, while $\ell$ does not)~\cite{Bouchbinder.14}. Consequently, material dependence (in addition to $v$-dependence) is expected to be encapsulated in the proportionality factor. To test this possibility, we should directly calculate $\ell$, instead of using its dimensional estimate. To calculate $\ell$ in our large-scale simulations, we need to estimate the actual length at which near tip nonlinearity becomes significant. Any energy functional $e_{\mbox{\tiny{strain}}}$ can be uniquely decomposed into its linear $e^{\mbox{\tiny{le}}}_{\mbox{\tiny{strain}}}$ and nonlinear $e^{\mbox{\tiny{nl}}}_{\mbox{\tiny{strain}}}\!\equiv\!e_{\mbox{\tiny{strain}}}-e^{\mbox{\tiny{le}}}_{\mbox{\tiny{strain}}}$ parts. The region near the tip at which $\|\partial_{_{\bm H}}e^{\mbox{\tiny{nl}}}_{\mbox{\tiny{strain}}}\|/\|\partial_{_{\bm H}}e^{\mbox{\tiny{le}}}_{\mbox{\tiny{strain}}}\|$ becomes non-negligible can be used to define $\ell$ ($\|\cdot\|$ is the magnitude of a tensor, i.e.~the square root of the sum of the squares of its elements). In particular, we define $\ell\!\equiv\!\sqrt{\cal A}$, where ${\cal A}$ is the area of the region in which $\|\partial_{_{\bm H}}e^{\mbox{\tiny{nl}}}_{\mbox{\tiny{strain}}}\|/\|\partial_{_{\bm H}}e^{\mbox{\tiny{le}}}_{\mbox{\tiny{strain}}}\|\!\ge\!0.5$ (see Fig.~S1 in~\cite{SM}), consistently with earlier definitions~\cite{Bouchbinder.14}. While the existence of a threshold (here $0.5$) involves some degree of arbitrariness in the definition of $\ell$, the results themselves do not strongly depend on the exact threshold (see Fig.~S2 in~\cite{SM}).

In Fig.~\ref{fig:wavelength}b we plot $\ell$ (as just defined) vs.~$\Gamma_0/\mu$ for both \mbox{\footnotesize NH} and \mbox{\footnotesize SVK} brittle materials. For both types of materials we observe $\ell\!\propto\!\Gamma_0/\mu$, as predicted theoretically. Moreover, the pre-factor in this relation is significantly larger for \mbox{\footnotesize SVK} materials compared to \mbox{\footnotesize NH} materials, which is consistent with the already discussed stronger nonlinearity of the former. In fact, the variability in the pre-factor appears to be similar in magnitude to the variability in the slope of the linear $\lambda\,$--$\,\Gamma_0/\mu$ relation shown in Fig.~\ref{fig:wavelength}a. To test whether the material dependence of the pre-factor indeed accounts for the material dependence of the wavelength shown in Fig.~\ref{fig:wavelength}a, we plot $\lambda$ vs.~$\ell$ in Fig.~\ref{fig:wavelength}c. We observe that the two curves approximately collapse one on top of the other, suggesting that in fact the oscillations wavelength follows a universal master curve $\lambda\!=\!\alpha\,\xi \!+\!\tilde\beta\,\ell$, where $\alpha$ and $\tilde\beta$ are nearly material independent. These results yet again strongly support the universal character of the 2D oscillatory instability.
\begin{figure}[!ht]
 \centering
 \includegraphics[width=0.48\textwidth]{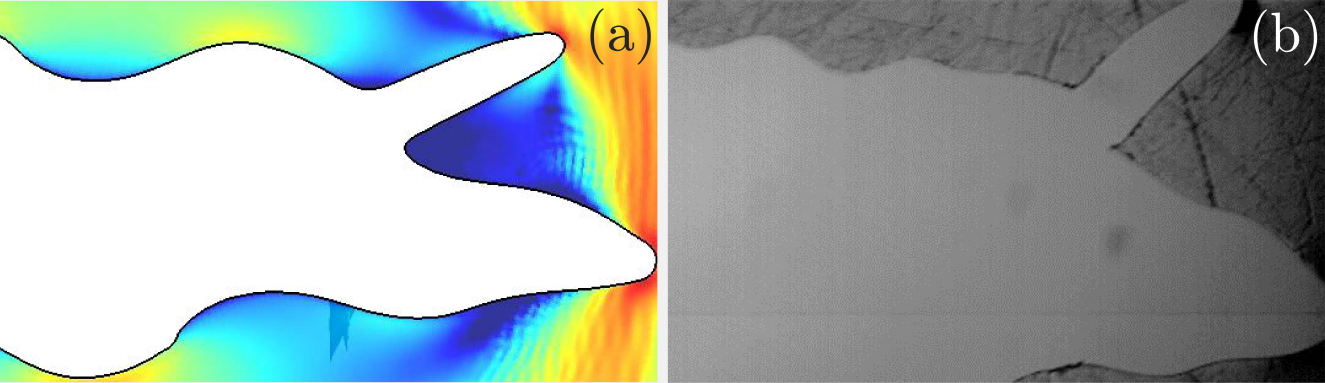}
 \caption{(a) Oscillations followed by tip-splitting in a phase-field simulation of \mbox{\footnotesize NH} materials under a large driving force (see Supplemental Movie S1). (b) The corresponding experimental observation (courtesy of Jay Fineberg, experimental details can be found in~\cite{boue2015origin}).}
 \label{fig:tip-splitting}
\end{figure}

When the driving force for fracture $W$, i.e.~the stored elastic energy per unit area ahead of the crack, is sufficiently large to allow cracks to accelerate to the critical velocity $v_c$, the oscillatory instability emerges. Are there additional, previously undiscovered instabilities in 2D dynamic fracture triggered when cracks are driven even more strongly? When $W$ is further increased, there exists a range of driving forces for which steady-state oscillatory cracks exist. Once a certain threshold value of $W$ is surpassed, a tip-splitting instability emerges, either after oscillations set in (asymmetric tip-splitting) or directly from the straight crack state (symmetric tip-splitting). This ultra-high-velocity tip-splitting instability in 2D dynamic fracture appears to be different from tip-splitting in 2D simulations~\cite{Abraham.94,Xu.94,Marder.95.jmps,Adda-Bedia.Ben-Amar.96,Adda-Bedia.99,Buehler.03,Adda-Bedia.04,Karma2004, Bouchbinder.05a,Spatschek.06, Pilipenko.07, Henry.08} and from the micro-branching instability in 3D experiments~\cite{Fineberg.99}. In Fig.~\ref{fig:tip-splitting}a we present tip-splitting that emerges after oscillations (see Supplemental Movie S1). Remarkably, such oscillations followed by tip-splitting have been recently observed under strong driving force conditions in previously unpublished experiments on brittle \mbox{\footnotesize NH} materials, see Fig.~\ref{fig:tip-splitting}b.
\begin{figure}[!ht]
 \centering
 \includegraphics[width=\columnwidth]{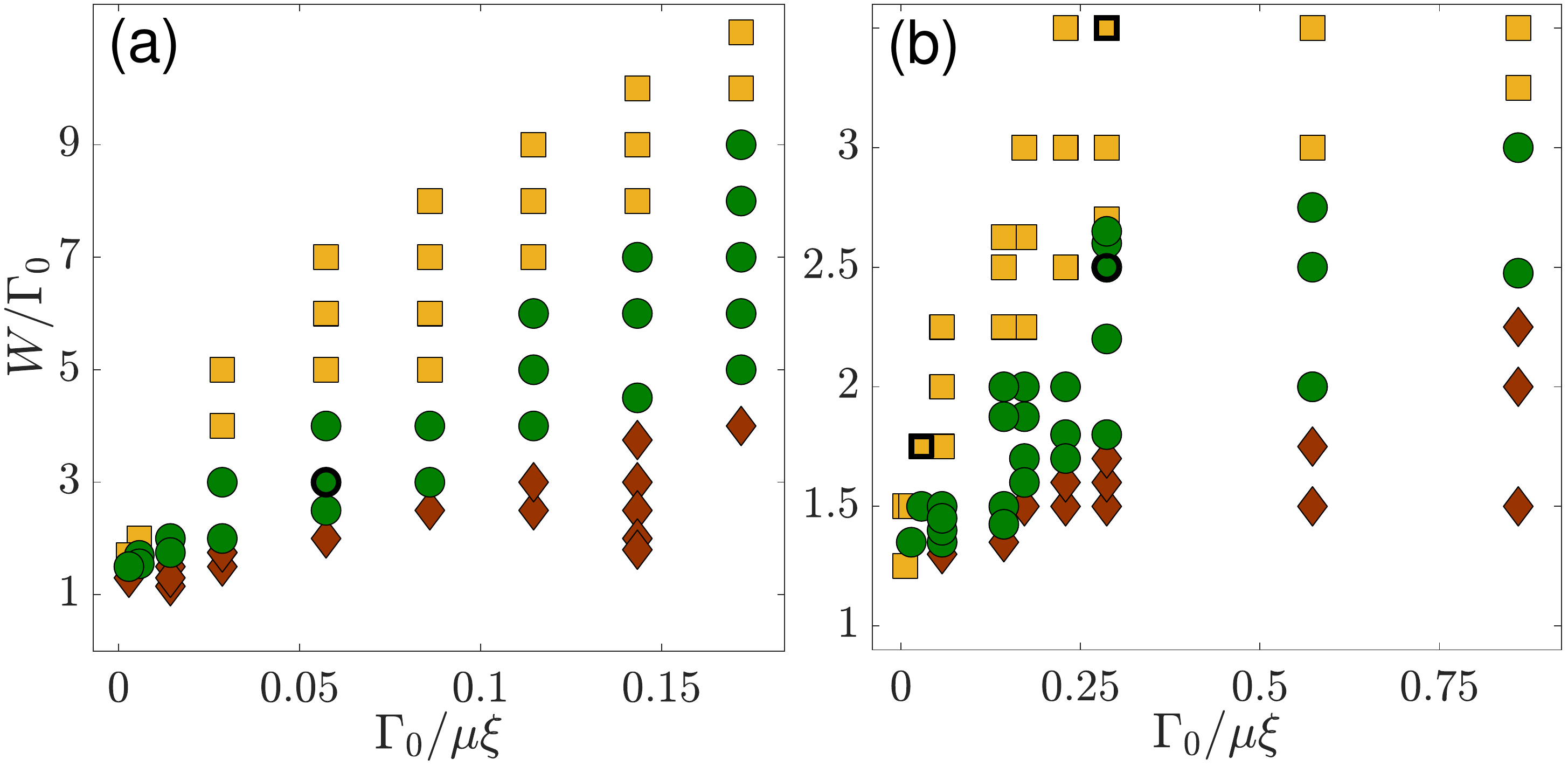}
 \caption{Stability phase-diagram of 2D dynamic fracture in the $W/\Gamma_0\,$--$\,\Gamma_0/\mu\xi$ plane for \mbox{\footnotesize SVK} (a) and \mbox{\footnotesize NH} materials (b). Shown are straight cracks (diamonds), oscillatory cracks (circles), and oscillatory or straight cracks undergoing tip-splitting (squares). See Supplemental Movies corresponding to the thick-lined squares in panel b. The upper thick-lined square also corresponds to Fig.~\ref{fig:tip-splitting}a. The thick-lined circles in panels a and b correspond to Figs.~\ref{fig:oscillations}b and~\ref{fig:oscillations}a, respectively.}
 \label{fig:phase-diagram}
\end{figure}

We are now in a position to construct a comprehensive stability phase-diagram of 2D dynamic fracture. The analysis presented above indicates that the relevant dimensionless parameters for such a stability phase-diagram are the ratio between the intrinsic nonlinear scale $\ell\!\propto\!\Gamma_0/\mu$ and the intrinsic dissipation scale $\xi$, and the ratio between the diving force for fracture $W$ and the fracture energy scale $\Gamma_0$. The stability phase-diagram in the $W/\Gamma_0\,$--$\,\Gamma_0/\mu\xi$ plane is shown in Fig.~\ref{fig:phase-diagram} for both brittle \mbox{\footnotesize SVK} (panel a) and \mbox{\footnotesize NH} (panel b) materials. For both classes of materials, for a fixed and finite $\Gamma_0/\mu\xi$, the phase-diagram exhibits the sequence of transitions with increasing $W/\Gamma_0$ described in the previous paragraph; for $W/\Gamma_0\!<\!1$ no crack propagation is possible (not shown), steady-state straight cracks exist upon increasing $W/\Gamma_0$ beyond unity over some range of $W/\Gamma_0$ (diamonds), then steady-state oscillatory cracks exist over some range for yet larger values of $W/\Gamma_0$ (circles), then tip-splitting emerges (squares), either after oscillations (asymmetric tip-splitting, see Supplemental Movie S1) or before (symmetric tip-splitting, see Supplemental Movie S2).

Interestingly, the $W/\Gamma_0$ range over which oscillatory cracks exist decreases with decreasing $\Gamma_0/\mu\xi$. In the limit $\Gamma_0/\mu\xi\!\to\!0$, it appears to vanish altogether such that straight cracks make a direct transition to tip-split cracks. While present numerical limitations do not allow us to conclude with certainty if the oscillatory instability exists or not in the absence of near tip elastic nonlinearity, they at least show that, if present, this instability would only exist over a vanishingly small range of loading, with a vanishingly small oscillations amplitude. Our results further strongly indicate that minute near tip nonlinearity, which is likely to exist in most materials, is sufficient to dramatically affect the instability whose wavelength essentially determined by the dissipation length $\xi$. Finally, and most remarkably, the salient properties and topology of the stability phase-diagram appear to be universal, i.e.~they are the same for two widely different forms of near tip nonlinearity.

In summary, we highlighted in this Letter the universal nature of the 2D oscillatory instability whose wavelength follows a universal master curve in terms of dissipation and nonlinear elastic intrinsic lengths, demonstrated the existence of a high-velocity tip-splitting instability and constructed a universal stability phase-diagram of 2D dynamic fracture. Future research should address the existence of the oscillatory instability in the absence of near tip elastic nonlinearity and the origin of the high-velocity tip-splitting instability.

\textit{Acknowledgments.---} This research was supported in part by the US-Israel Binational Science Foundation (BSF) grant
no.~2012061, which provided partial support for C.-H.C. E.B.~acknowledges support from
the William Z.~and Eda Bess Novick Young Scientist Fund and the Harold Perlman Family. A.K.~acknowledges support of grant number DE-FG02-07ER46400 from the US
Department of Energy, Office of Basic Energy Sciences. We are grateful to J.~Fineberg for providing the experimental result shown in Fig.~\ref{fig:tip-splitting}b.

%

\newpage
\onecolumngrid
\begin{center}
	\textbf{\large Supplemental Material for: ``Universality and stability phase-diagram of two-dimensional brittle fracture''}
\end{center}

\setcounter{equation}{0}
\setcounter{figure}{0}
\setcounter{section}{0}
\setcounter{table}{0}
\setcounter{page}{1}
\makeatletter
\renewcommand{\theequation}{S\arabic{equation}}
\renewcommand{\thefigure}{S\arabic{figure}}
\renewcommand{\thesubsection}{S-\Roman{subsection}}
\renewcommand*{\thepage}{S\arabic{page}}
\renewcommand{\bibnumfmt}[1]{[S#1]}
\renewcommand{\citenumfont}[1]{S#1}
\twocolumngrid

In this Supplemental Material file we provide additional technical information about the calculations presented in the manuscript.

\subsection{Phase-field theory, parameters and\\ numerical implementation}

The phase-field theory of fracture used in the manuscript is presented in [C.-H.~Chen, E.~Bouchbinder, and A.~Karma, Nature
Physics {\bf 13}, 1186 (2017)]. We briefly repeat it here for completeness.  The theory formulated in terms of the Lagrangian
$L\!=\!T-U$, where $U\!=\!\int\left[\tfrac{1}{2}\kappa\left(\nabla\phi\right){}^{2}+g\left(\phi\right)\left(e_{\mbox{\tiny{strain}}}-e_{c}\right)\right]dV$
is the potential energy and $T\=\tfrac{1}{2}\rho\!\int\!g(\phi)(\partial_{t}\bm{u})^{2}dV$
is the kinetic energy, $\rho$ is the mass density and $dV$ is a volume
element. $e_{\mbox{\tiny{strain}}}\left(\bm{H}\right)$
is the elastic strain energy functional, where $\bm{H}=\nabla\bm{u}$
is the displacement gradient tensor and $\bm{u}\left(x,y,t\right)$
is the displacement vector field. In this study we considered tensile (mode-I) fracture where cracks propagate in the $x$-direction and the tensile loading is applied in the $y$-direction.

The scalar phase-field $\phi\left(x,y,t\right)$ varies smoothly in space between the pristine ($\phi\!=\!1$) and
the fully broken ($\phi\!=\!0$) states of the material. The broken
state becomes energetically favored when $e_{\mbox{\tiny{strain}}}$
exceeds a threshold $e_{c}$, and the function $g\left(\phi\right)=4\phi^{3}-3\phi^{4}$
decreases with decreasing $\phi$ such that the elastic constants
vanish as $\phi\!\to\!0$ at large strains, which exactly corresponds
to the traction-free boundary conditions that define a crack. Spatial
gradients of $\phi$ are energetically penalized, as quantified by
$\tfrac{1}{2}\kappa\left(\nabla\phi\right){}^{2}$. Finally, the evolution
equations for $\phi$ and $\bm{u}$ are derived from Lagrange's equations
$\ensuremath{\partial_{t}\left[\delta L/\delta\left(\partial_{t}\psi_{i}\right)\right]-\delta L/\delta\psi_{i}+\delta D/\delta\left(\partial_{t}\psi_{i}\right)\=0}$
for $\bm{\psi}=\left(\phi,\bm{u}\right)$, where the functional $D=(2\chi)^{-1}\!\int(\partial_{t}\phi)^{2}dV$
controls the rate of energy dissipation. The latter features a dissipation-related
length scale $\xi\!=\!\sqrt{\kappa/\left(2e_{c}\right)}$ and a fracture
energy $\Gamma\left(v\right)$, the energy dissipated per unit crack
surface when the crack propagation is $v$.

The quasi-static the fracture energy is given by $\Gamma_0\!\equiv\!\Gamma\left(v\!\to\!0\right)\!=\!2\sqrt{2\kappa e_{c}}\int_{0}^{1}d\phi\sqrt{1-g(\phi)}\!\approx\!2.866e_c\xi$, which implies that the normalized nonlinear elastic length $\ell/\xi\!\propto\!\Gamma_0/\mu\xi\!\approx\!2.866e_c/\mu$ can be varied by varying the dimensionless ratio $e_c/\mu$. Our systems are loaded in mode-I by fixing $u_y$ at the horizontal boundaries by a small amount corresponding to strains of up to a few percent, ensuring that the system remains in the linear elastic regime everywhere except for a small region near the crack tip. The loading can be quantified by the normalized stored elastic energy per unit area far ahead of the crack, $W/\Gamma_0$. Our systems are chosen to be effectively large enough to reduce the finite size effects. In particular, in order to study long crack propagation distances, the crack tip in the simulations is kept at the center of simulation box by using a treadmill approach, which removes a layer at the left vertical boundary behind the crack and adds a strained layer at the right vertical boundary ahead of it. The loading and system sizes used in this study are presented in Table~\ref{tab:System-parameters-Fig.} and in its caption.

The numerical implementation details, including spatial grid, time spacing and convergence tests,
are identical to those of [C.-H.~Chen, E.~Bouchbinder, and A.~Karma, Nature Physics {\bf 13}, 1186 (2017)].
The phase-field parameters that have not been varied as detailed above, have been set identical to those of [C.-H.~Chen, E.~Bouchbinder, and A.~Karma, Nature
Physics {\bf 13}, 1186 (2017)]. For the \mbox{\footnotesize SVK} model, which has not been studied before, we have set $\tilde{\lambda}\=2\mu$ such that its linear approximation identifies with that of the \mbox{\footnotesize NH} energy functional.

\begin{table}[h]
\begin{centering}
\begin{tabular}{|c|c|c|c|}
\hline
Figure & $\Gamma_0/\mu\xi\!\approx\!2.866\!\times\!e_{c}/\mu$ & $W/\Gamma_{0}$ & System size\tabularnewline
\hline
\hline
1a & $2.866\times0.1$ & $2.5$ & $300\times300\,\xi^{2}$\tabularnewline
\hline
1b & $2.866\times0.02$ & $3.0$ & $200\times200\,\xi^{2}$\tabularnewline
\hline
1c \mbox{\footnotesize NH} & $2.866\times0.005$ & $1.35$ & $150\times150\,\xi^{2}$\tabularnewline
\hline
1c \mbox{\footnotesize SVK} & $2.866\times0.005$ & $2.0$ & $200\times200\,\xi^{2}$\tabularnewline
\hline
3a & $2.866\times0.1$ & $3.5$ & $400\times400\,\xi^{2}$\tabularnewline
\hline
\end{tabular}
\par\end{centering}
\caption{Loading and system sizes (length $\times$ width) used in Fig.~1a-c and Fig.~3a in the manuscript. In Fig.~2 and Fig.~4 in the manuscript system sizes range from $150\times150\,\xi^{2}$ to $600\times600\,\xi^{2}$, and the applied strains range for $0.1\%$ to $7\%$.\label{tab:System-parameters-Fig.}}
\end{table}

\begin{figure*}[ht!]
\begin{centering}
\includegraphics[scale=0.44]{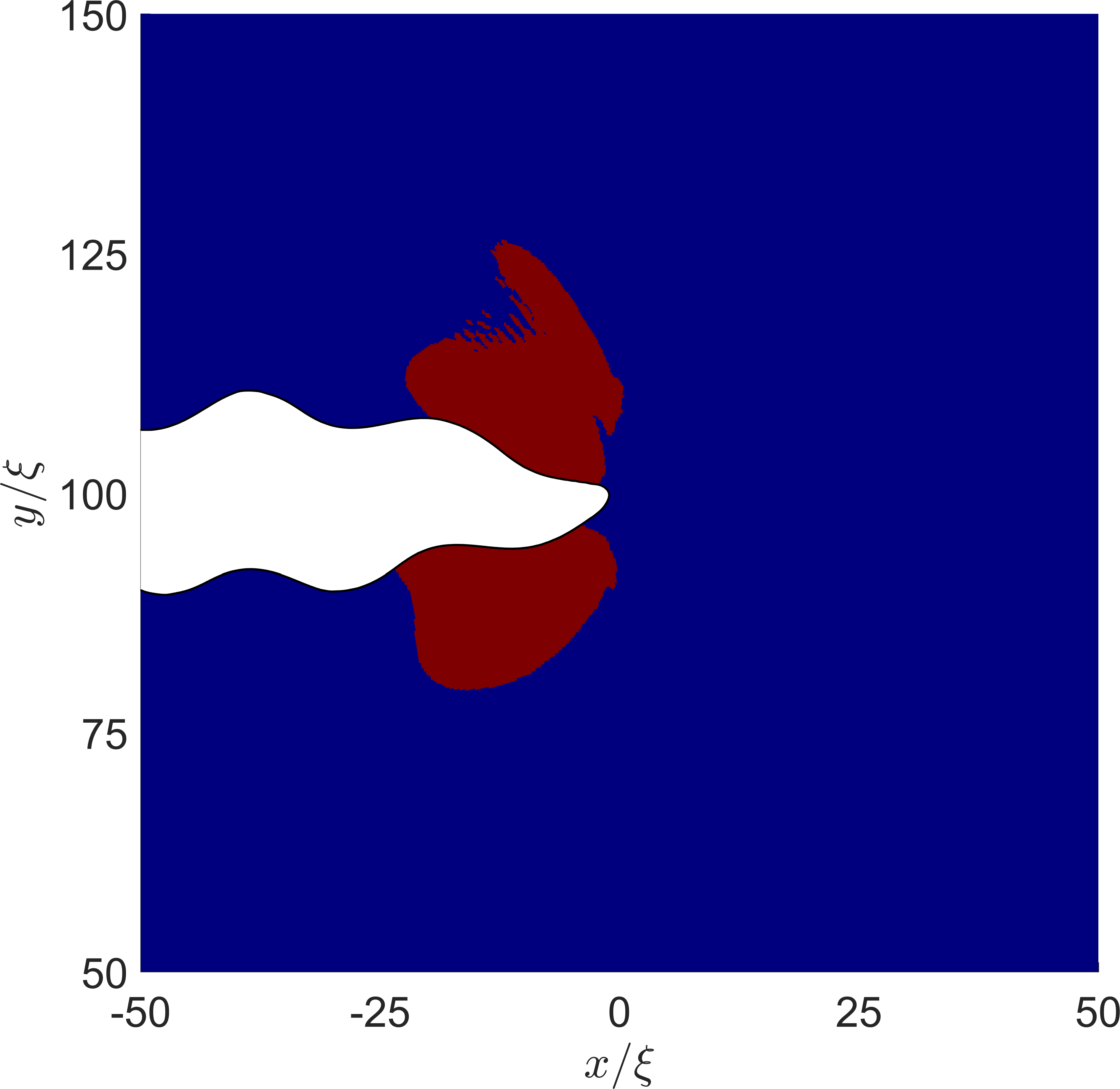}
\end{centering}
\begin{centering}
\includegraphics[scale=0.44]{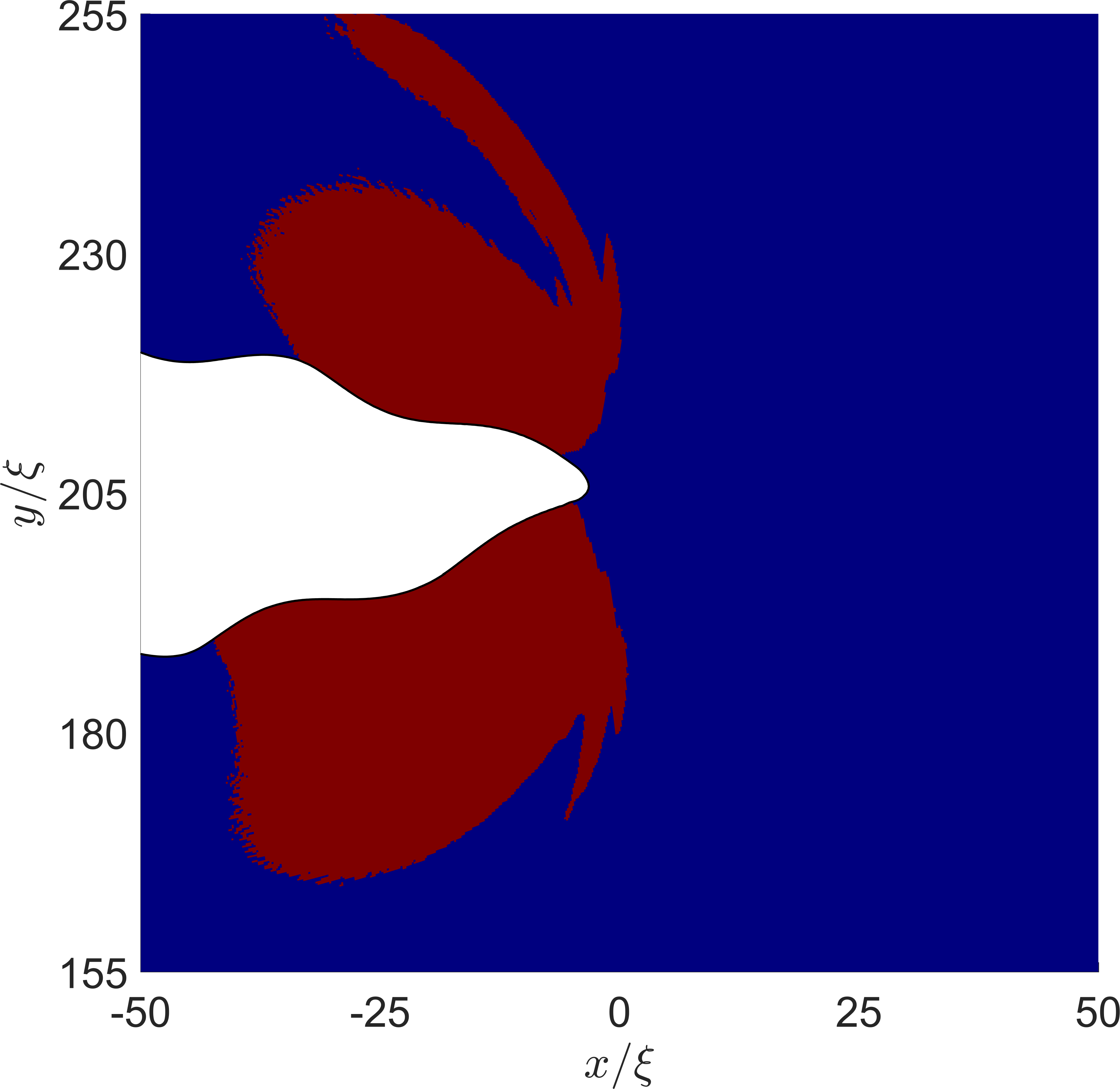}
\par\end{centering}
\caption{The region that satisfies $\|\partial_{_{\bm H}}e^{\mbox{\tiny{nl}}}_{\mbox{\tiny{strain}}}\|/\|\partial_{_{\bm H}}e^{\mbox{\tiny{le}}}_{\mbox{\tiny{strain}}}\|\!\ge\!0.5$ (red) for the \mbox{\footnotesize NH} energy
functional for (left) $e_{c}/\mu=0.1$, $W/\Gamma_{0}=2.5$ and system size of $200\times200\,\xi^{2}$, and for (right)
$e_{c}/\mu=0.2$, $W/\Gamma_{0}=2.5$  and system size of $400\times400\,\xi^{2}$. That is, both the dimensionless loading $W/\Gamma_{0}$ and applied strain have been fixed, while $e_{c}/\mu$ has been varied.
The $2$-fold variation in $e_c/\mu$ in this example resulted in a variation of the area $\mathcal{A}$ of the red region between the two cases that corresponds to a $1.86$-fold variation in the nonlinear length $\ell$ (close to the predicted $2$-fold variation). Similar results are obtained for the \mbox{\footnotesize SVK} energy functional.
\label{fig.ell_shape}}
\end{figure*}

\subsection{The truncated neo-Hookean energy functional and its regularization}

The \mbox{\footnotesize NH} energy functional truncated to $4^{\rm th}$ order in $\bm{H}$ is given in Eq.~(2) in the manuscript. Formally, the leading order nonlinearity include terms up to $3^{\rm rd}$ order in $\bm{H}$. The \mbox{\footnotesize NH} energy functional truncated to $3^{\rm rd}$ order in $\bm{H}$ is not bounded from below (in particular, it is not bounded from below under uniaxial tension or volumetric expansion) and hence a global fracture problem defined by it is not well-posed. Consequently, we considered the $4^{\rm th}$ order truncation scheme.
\begin{figure}[ht!]
\begin{centering}
\includegraphics[scale=0.47]{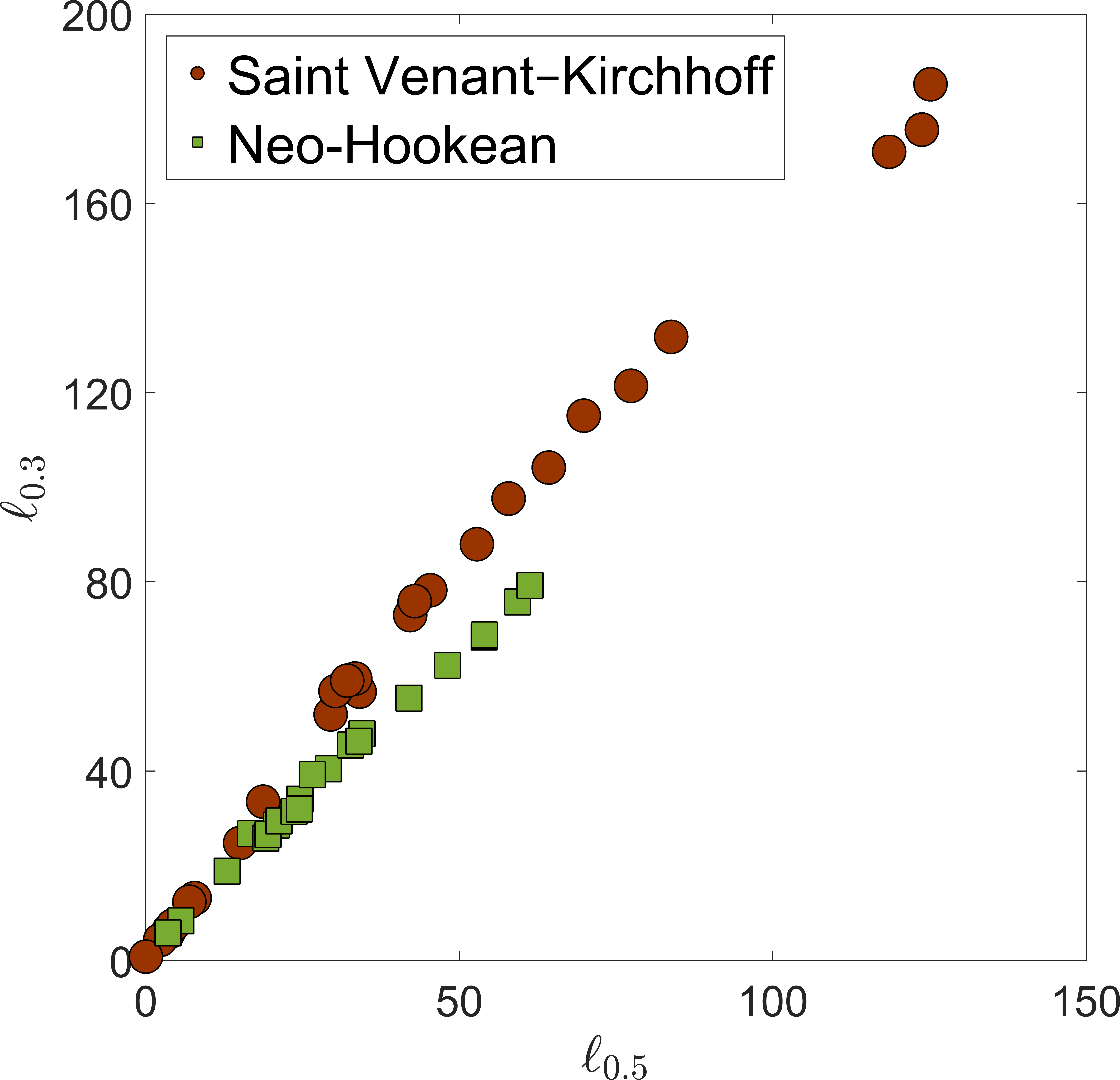}.
\par\end{centering}
\caption{The nonlinear elastic length $\ell$ with a $0.3$ threshold, $\ell_{0.3}$ vs.~$\ell$ with a $0.5$ threshold, $\ell_{0.5}$ (as in the manuscript), for
the \mbox{\footnotesize SVK} (circles) and \mbox{\footnotesize NH} (squares) energy functionals. $\ell_{0.3}$ is proportional to $\ell_{0.5}$, with a proportionality factor which is essentially identical for the two widely different energy functionals. \label{fig:-with-threshold}}
\end{figure}

In fact, while the $4^{{\rm th}}$ order truncated $\mbox{\footnotesize NH}$ energy functional is bounded from below, it still features a nonphysical minimum with energy lower than that of the $\boldsymbol{H}\=0$ state. Consequently, it should be regularized once the derived dynamics are implemented. Our regularization scheme is based on the following observation: any physical energy functional must satisfy the condition $e_{\mbox{\tiny{strain}}}\left(\left(1+\delta\right)\boldsymbol{H}\right)\!\geq\! e_{\mbox{\tiny{strain}}}\left(\boldsymbol{H}\right),\forall\delta\!\ge\!0$, or equivalently  $\left.\partial_{\delta}e_{\mbox{\tiny{strain}}}\left(\left(1+\delta\right)\boldsymbol{H}\right)\right|_{\delta=0}\!\ge\!0$, i.e the elastic strain energy is a non-decreasing function of deformation. Whenever this condition is violated, the dynamics are regularized by replacing the Piola stress, $\partial_{\boldsymbol{H}}e_{\mbox{\tiny{strain}}}$, by its linear approximation, $\partial_{\boldsymbol{H}}e_{\mbox{\tiny{strain}}}^{\mbox{\tiny le}}$ (as defined in the manuscript), which always satisfies the non-decreasing condition. Whenever the regularization is activated, there is a local discontinuity in the stress; yet, our results show that the regularization is activated in compact and sparse regions, away from the crack tip, hence having minor influence on the dynamics.

\subsection{The nonlinear elastic length $\ell$}

In order to quantitatively extract the nonlinear elastic length $\ell$ from our simulations, we defined in the manuscript the region around the crack tip that satisfies $\|\partial_{_{\bm H}}e^{\mbox{\tiny{nl}}}_{\mbox{\tiny{strain}}}\|/\|\partial_{_{\bm H}}e^{\mbox{\tiny{le}}}_{\mbox{\tiny{strain}}}\|\!\ge\!0.5$. This region is shown
in Fig.~\ref{fig.ell_shape} for \mbox{\footnotesize NH} materials for two values of $e_c/\mu$ separated by a factor of $2$; indeed, the corresponding regions feature different sizes (see figure caption for details). The area of these regions ${\cal A}$ is then used to define the nonlinear elastic length through $\ell\equiv\sqrt{\mathcal{A}}$, which has been shown to satisfy $\ell\!\propto\!\Gamma_0/\mu$, as predicted theoretically (Fig.~2b in the manuscript).

The definition of $\ell$ involves a threshold ($0.5$ in the manuscript), whose exact value is somewhat arbitrary. In Fig.~\ref{fig:-with-threshold} we show that the exact value of the threshold has no effect on our results. In particular, we replaced the threshold $0.5$ in the definition above by $0.3$ for both the \mbox{\footnotesize NH} and \mbox{\footnotesize SVK} energy functionals, and showed that this difference only affects the magnitude of the proportionality factor in the relation $\ell\!\propto\!\Gamma_0/\mu$, which is essentially identical for the two widely different energy functionals, thus having no real effect on our results.

\subsection{Description of the supplemental movies}

We attach to this Supplemental Material two movies which present a tip-splitting instability (corresponding to the thick-lined squares in Fig.~4b for $\mbox{\footnotesize NH}$ materials). These movies are shown in a co-moving reference frame in which the crack always remains in the center of the box (the instantaneous magnitude of the single tip velocity is shown up to the tip-splitting event, and the dimensionless time is presented as well). Brief movie descriptions:
\begin{itemize}[leftmargin=*]
\item \textbf{Movie\_S1:} A rapid oscillatory crack undergoing an asymmetric tip-splitting instability (the movie corresponds to the upper thick-lined square in Fig.~4b in the manuscript). Notice the attempted tip-splitting events prior to the actual tip-splitting instability. A snapshot of the post instability state is shown in Fig.~3a in the manuscript (where the color code represents $\log(e_{\mbox{\tiny{strain}}}/e_c)$, rather than $e_{\mbox{\tiny{strain}}}/e_c$ itself).
\item \textbf{Movie\_S2:} A rapid straight crack undergoing a symmetric tip-splitting instability (the movie corresponds to the lower thick-lined square in Fig.~4b in the manuscript).
\end{itemize}
The videos are available \href{http://www.weizmann.ac.il/chemphys/bouchbinder/universality-and-stability}{here}.

\end{document}